\documentclass[twocolumn]{aa}
\usepackage{graphicx}
\begin{document}
   \title{Low-column density gas clumps in the halo of the Milky Way
          \thanks{\small Based on observations carried out at the European
          Southern Observatory (ESO), La Silla, under prog.\,ID No.\,
          166.A-0106 with the UVES spectrograph at the ESO Very Large
	  Telescope, Paranal, Chile.}}

   \author{Philipp Richter,
          \inst{1}
          Tobias Westmeier,
          \inst{2}
          \and
          Christian Br\"uns
          \inst{2}
          }


   \institute{
              Institut f\"ur Astrophysik und Extraterrestrische
              Forschung, Universit\"at Bonn, \\
              Auf dem H\"ugel 71, 53121 Bonn, Germany; prichter@astro.uni-bonn.de
         \and
              Radioastronomisches Institut, 
              Universit\"at Bonn, 
              Auf dem H\"ugel 71, 53121 Bonn, Germany
            }

   \date{Received xxx, 2005; accepted xxx}

\abstract{
We report on the detection of low-column density neutral hydrogen
clumps in the halo of the Milky Way. Using high-resolution 
(FWHM$\sim7$ km\,s$^{-1}$) optical spectra obtained with the VLT/UVES
spectrograph we detect narrow interstellar absorption by 
Ca\,{\sc ii} and Na\,{\sc i} at high radial velocities near $v_{\rm LSR}
\approx -150$ km\,s$^{-1}$ toward the quasar PKS\,1448$-$232
($l=335.4, b=+31.7$). Follow-up H\,{\sc i} 21cm observations
with the VLA unveil the presence of a complex of small neutral
hydrogen clumps with H\,{\sc i} column densities $< 8 \times
10^{18}$ cm$^{-2}$. The measured H\,{\sc i} line widths imply
that the gas is relatively cold with temperatures $T\leq900$ K. Although
the line of sight towards PKS\,1448$-$232 does not pass
immediately through a known large high-velocity cloud (HVC), 
the sky position and the measured radial velocities 
suggest that these clumps are associated with HVC
cloud complex L. An inspection of other UVES quasar spectra
shows that weak, narrow Ca\,{\sc ii} 
absorption at high velocities is a common phenomenon, even in
directions where high-velocity H\,{\sc i} 21cm emission is not detected.
This suggests that the Milky Way halo contains a large number
of high-velocity neutral gas clumps with low H\,{\sc i} column densities.
If such clumps are typical for halos of spiral galaxies, they 
should contribute significantly to the population of strong
Mg\,{\sc ii} absorbers and Lyman-Limit Systems (LLS) seen
in the circumgalactic environment of other galaxies.

   }
   \titlerunning{Gas clumps in the halo of the Milky Way
}
   \maketitle
%

\section{Introduction}

During the last decades, absorption and emission line measurements 
have demonstrated that the Milky Way is surrounded by a complex, multi-phase
gaseous halo. Embedded in a corona of million-degree gas, neutral and
ionized gas clouds move with high radial velocities ($>40$ km\,s$^{-1}$) through
the Milky Way halo, giving rise to the population of intermediate- 
and high-velocity clouds (IVCs and HVCs, respectively; Wakker 2004).
The origin of these clouds (Galactic or extragalactic) was uncertain for many
years. However, recent abundance measurements have shown that the 
Galaxy is accreting gas from satellite galaxies and from 
intergalactic space, but at the same time is circulating gas from the disk
into the halo by way of a galactic fountain 
(see Bregman 2004 for a review).
While most of the recent HVC measurements have focussed on
the distribution and properties of the large IVC and HVC complexes (e.g.,
Wakker et al.\,1999; Richter et al.\,2001), compact high-velocity clouds
(CHVCs; e.g., Braun \& Burton 1999; Westmeier et al.\,2005), and 
highly-ionized HVCs (e.g., Sembach et al.\,2003),
relatively little attention has been paid so far to the abundance and
distribution of neutral gaseous structures in the halo that have H\,{\sc i} column densities
below the detection limit of the large 21cm HVC surveys ($\sim 10^{19}$
cm$^{-2}$). 
Using more sensitive Arecibo 21cm data, Hoffman et al.\,(2004) have 
found a population of ``mini-HVCs'' with low H\,{\sc i} column
densities (a few $10^{18}$ cm$^{-2}$, typically) and small
angular diameters ($\leq 35'$).
Towards the Large Magellanic Cloud, Richter et al.\,(2003) have
detected FUV absorption by molecular hydrogen and various weakly ionized metals
in a dense gas filament in the halo that has a total H\,{\sc i} column
density of $\sim 10^{18}$ cm\,$^{-2}$ and a thickness of only 
$\sim 40$ AU.

In this paper we discuss low-column density, small-scale structures
in the halo that have been detected in Ca\,{\sc ii} and Na\,{\sc i} absorption 
toward the quasar (QSO) PKS\,1448$-$232 and other extragalactic background sources. Studying the
frequency and distribution of these low-column density absorbers is important
to better understand the connection between the different gas phases
in the Milky Way halo, and to link the column density distribution
of neutral gas in the halo to the properties of absorption line systems 
in the circumgalactic environment of other galaxies.

\section{Observations and data handling}

The optical spectral data for the quasar PKS\,1448$-$232 
($l=335.4$, $b=+31.7$, $z_{\rm em}=2.2$) were obtained 
in June 2001 with the {\it Ultraviolet and Visual Echelle Spectrograph} 
(UVES) at the ESO Very Large Telescope as part of the UVES Large Programme 
{\it The Evolution of the Intergalactic Medium} (PI: J. Bergeron).
The data provide a spectral resolution of $R\sim 42000$, corresponding
to $6.6$ km\,s$^{-1}$ FWHM. 
The raw data were reduced using the UVES pipeline
implemented in the ESO-MIDAS software package.
The pipeline reduction includes flat-fielding,
bias- and sky-subtraction, and a relative
wavelength calibration. 
The signal-to-noise (S/N) per resolution
element is $\sim 110$ in the region of the Ca\,{\sc ii} absorption 
near $3950$ \AA, and $\sim 190$ near the Na\,{\sc i} lines ($\sim 5895$ \AA). The spectra
were analyzed using the FITLYMAN program in MIDAS (Fontana \& Ballester 1995),
which delivers velocity centroids\footnote {Radial velocities cited
in this paper refer to the Local 
Standard of Rest (LSR).},
column densities and Doppler parameters ($b$ values)
from Voigt-profile fitting. 

The follow-up H\,{\sc i} 21cm observations (two coverages of 6~hours each) were carried 
out in June 2004 with the Very Large Array (VLA) in the DnC 
configuration, which is especially well suited for sources with a 
declination of $\delta < -15^{\circ}$. We chose 1331$+$305 (3C286) as 
flux and bandpass calibrator and 1526$-$138 as gain calibrator. For each 
of the two polarisations the correlator provided a bandwidth of 1.56~MHz 
with 256 spectral channels, resulting in a channel separation of $1.3 \; 
{\rm km \, s}^{-1}$. After flagging the data affected by radio frequency 
interference, we carried out the standard bandpass, flux and gain 
calibration with AIPS. To further improve the gain calibration, we 
self-calibrated on the continuum sources in an iterative procedure. We 
then deconvolved the data cube using the CLEAN algorithm 
(originally developped by H\"ogbom 1974). To increase the 
signal-to-noise ratio, a Gaussian $uv$ taper with a radius of $1 \; {\rm 
k} \lambda$ at the 30\% level was applied. In addition, we smoothed the 
velocity resolution to $5.1 \; {\rm km \, s}^{-1}$. The resulting 
synthesised beam has a FWHM of about $2'$. The rms in the final data 
cube is $1.4 \; {\rm mJy/beam}$ towards the centre of the field, 
corresponding to a brightness temperature rms of $60 \; {\rm mK}$.

\begin{figure}[t!]
\resizebox{0.9\hsize}{!}{\includegraphics{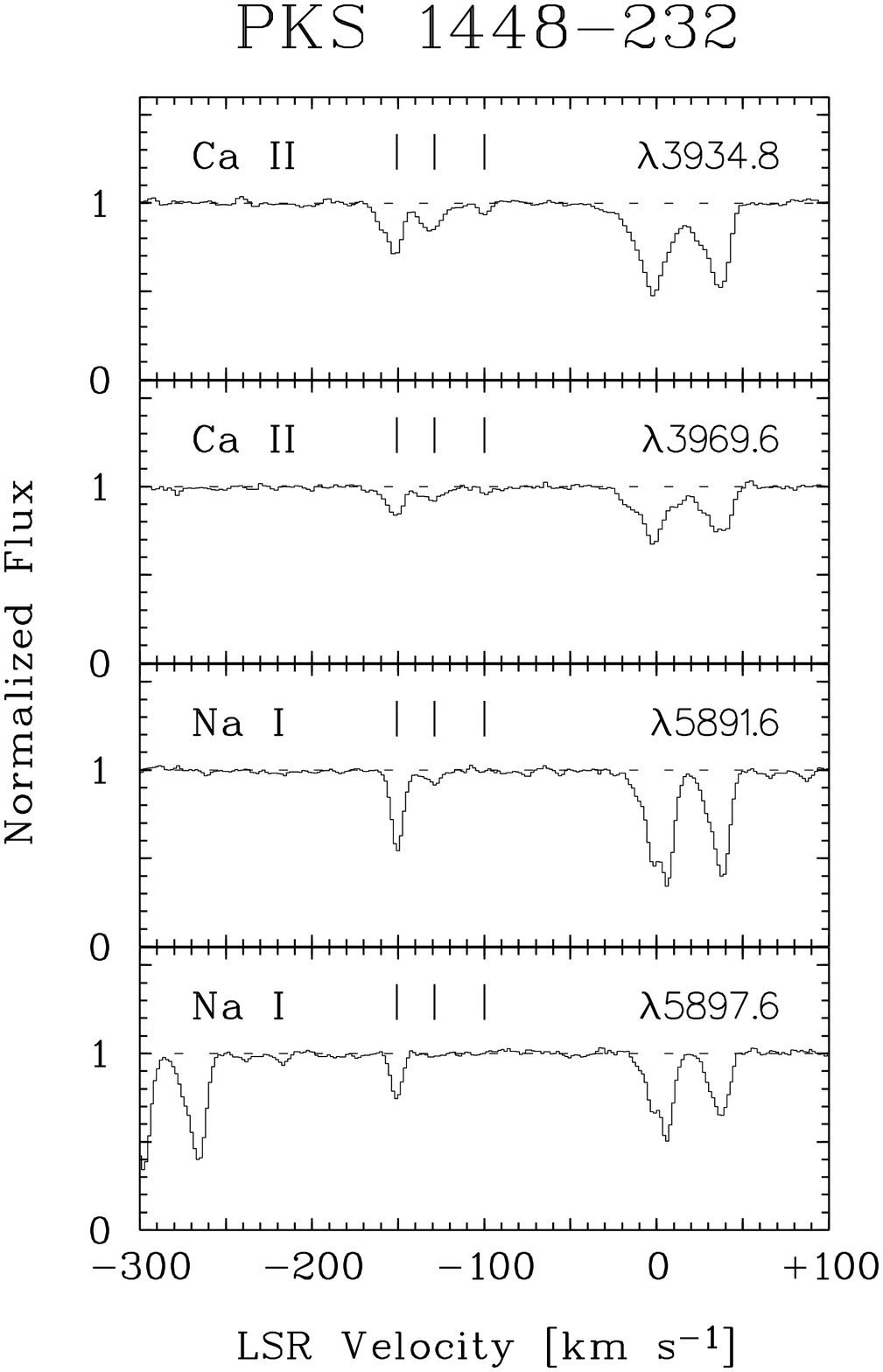}}
\caption[]{
Ca\,{\sc ii} and Na\,{\sc i} absorption profiles in the UVES spectrum of
PKS\,1448$-$232. Three high-velocity absorption components 
related to gas in the Galactic halo
are seen
at $-101,-130$ and $-152$ km\,s$^{-1}$ (indicated with the long dashes).
}
\end{figure}

\begin{table}[t!]
\caption[]{Summary of UVES Ca\,{\sc ii} and Na\,{\sc i} measurements toward PKS\,1448$-$232}
\begin{footnotesize}
\begin{tabular}{lllll}
\hline
$v_{\rm LSR}$  & log $N_{\rm Ca\,II}$ & $b_{\rm Ca\,II}$ & log $N_{\rm Na\,I}$ 
& $b_{\rm Na\,I}$ \\
$[$km\,s$^{-1}$] & [$N$ in cm$^{-2}$]     & [km\,s$^{-1}$]   &   [$N$ in cm$^{-2}$]       & [km\,s$^{-1}$] \\
\hline
$-101$         & $11.00\pm0.04$       & $4.0\pm1.2$      & $\leq 10.84$        &   ...            \\
$-130$         & $11.62\pm0.03$       & $8.3\pm2.7$      & $10.96\pm0.10$      & $5.1\pm2.1$      \\
$-152$         & $11.83\pm0.03$       & $6.6\pm2.3$      & $11.77\pm0.04$      & $2.2\pm1.3$      \\
\hline
\end{tabular}
\end{footnotesize}
\end{table}

\section{Results}

Fig.\,1 shows the absorption profiles of Ca\,{\sc ii} 
$\lambda 3934.8$, Ca\,{\sc ii} $\lambda 3969.6$,
Na\,{\sc i} $\lambda 5891.6$, and Na\,{\sc i} $\lambda 
5897.6$ from the UVES spectrum of 
PKS\,1448$-$232 plotted on an LSR velocity scale. Next to the absorption components near
zero velocity that refer to gas in the local Galactic disk, high-velocity Ca\,{\sc ii} 
and Na\,{\sc i} absorption from gas in the halo 
is visible in the range $-100$ to $-150$ km\,s$^{-1}$. 
For the high-velocity gas, three individual velocity components at 
$-101,-130$ and $-152$ km\,s$^{-1}$ can be identified. Logarithmic Ca\,{\sc ii} and Na\,{\sc i} 
column densities for these components range between $\sim 10.8$ and $\sim 11.8$. Measured 
column densities and $b$ values are summarized in Table 1. 

No high-velocity H\,{\sc i} 21cm emission at this position is seen in the H\,{\sc i} data of the 
Leiden/Argentine/Bonn (LAB) survey (Kalberla et al.\,2005; Hartmann \& Burton 1997), but our
follow-up high-resolution VLA observations unveil the presence of several high-velocity H\,{\sc i} clumps
at low H\,{\sc i} column densities (Fig.\,2)\footnote{Note that follow-up
Effelsberg H\,{\sc i} 21cm observations
confirm the presence of these H\,{\sc i} structures.}.
Four of these clumps
are labeled with the letters A, B, C, and D (Fig.\,2). These clumps have typical peak-brightness
temperatures of $T_{\rm B} \sim 0.3-0.4$ K, corresponding to H\,{\sc i} column densities of 
$\sim 4-8 \times 10^{18}$ cm$^{-2}$. Measurements of these features are summarized in Table 2.
As indicated, the sight line towards PKS\,1448$-$232 passes the outer envelope of clump A.
The mean LSR velocity
of the H\,{\sc i} emission is approximately $-150$ km\,s$^{-1}$, thus similar to what is found 
for the strongest component of the Ca\,{\sc ii} and Na\,{\sc i} absorption. No significant H\,{\sc i} emission
is seen at velocities $\geq -140$ km\,s$^{-1}$, implying that the H\,{\sc i} column densities
of the Ca\,{\sc ii} absorption components at $-101$ and $-130$ km\,s$^{-1}$ fall below the 
detection limit of the 21cm observations.
Velocity sub-structure is present in clumps A, B, and D, while the profile of C is consistent with 
a single Gaussian component. From the velocity width of the H\,{\sc i} emission in clump C we 
obtain an upper limit for the kinetic gas temperature of $T_{\rm kin}\leq 900$ K. However, in view
of the relatively large Na\,{\sc i}/Ca\,{\sc ii} ratio and the possible presence of non-thermal gas motions,
the actual gas temperature could be much lower than that. 

Although the column densities for both Ca\,{\sc ii} and Na\,{\sc i} are measured with 
high accuracy, unknown ionization
conditions and dust depletion effects as well as the limited beam size of the H\,{\sc i} 
measurements makes it impossible to reliably measure the metal abundance in the gas
(see also Wakker 2001). 
Given the sky position of the clumps and the observed
velocity range, it is likely that the gas is associated with HVC complex L, which is only 
a few degrees away (Wakker et al.\,2004). If so, our Ca\,{\sc ii} and Na\,{\sc i} 
measurements represent the first secure detection of this HVC in absorption. Weiner et al.\,(2001)
and Putman et al.\,(2003) have observed strong H\,$\alpha$ emission from complex L with 
typical fluxes of several hundred mR. 
According to their models of the Galactic ionizing flux, this places
complex L in the lower Galactic halo with a distance of $\sim 8-22$ kpc and a $z$ height 
of $\sim 4-12$ kpc above the Galactic center region. 
From the typical angular size of the H\,{\sc i} clumps ($\sim 3'$)
and a typical H\,{\sc i} column density of $N$(H\,{\sc i}$)=5\times10^{18}$ cm$^{-2}$
we can infer an estimate of the gas density 
as a function of distance via      
$n$(H\,{\sc i})/cm$^{-3}=1.85\,(d/$kpc)$^{-1}$ (note that for this estimate we 
assume spherical symmetry, which may not be a good approximation for the true
shape of the cloud). For $d=4-12$ kpc we obtain
$n$(H\,{\sc i}$)\approx 0.1-0.2$ cm$^{-3}$. This, together with $T_{\rm kin}\leq 900$ K
from the H\,{\sc i} line widths, implies that the thermal pressure is 
$P/k = nT \leq 180$ cm$^{-3}$\,K. This pressure is relatively low 
but is consistent with values predicted for high-velocity galactic fountain
gas at $z\sim10$ kpc with solar abundances and moderate dust destruction
(Wolfire et al.\,1995; their Fig.\,1).

\begin{figure}[t!]
\resizebox{0.92\hsize}{!}{\includegraphics{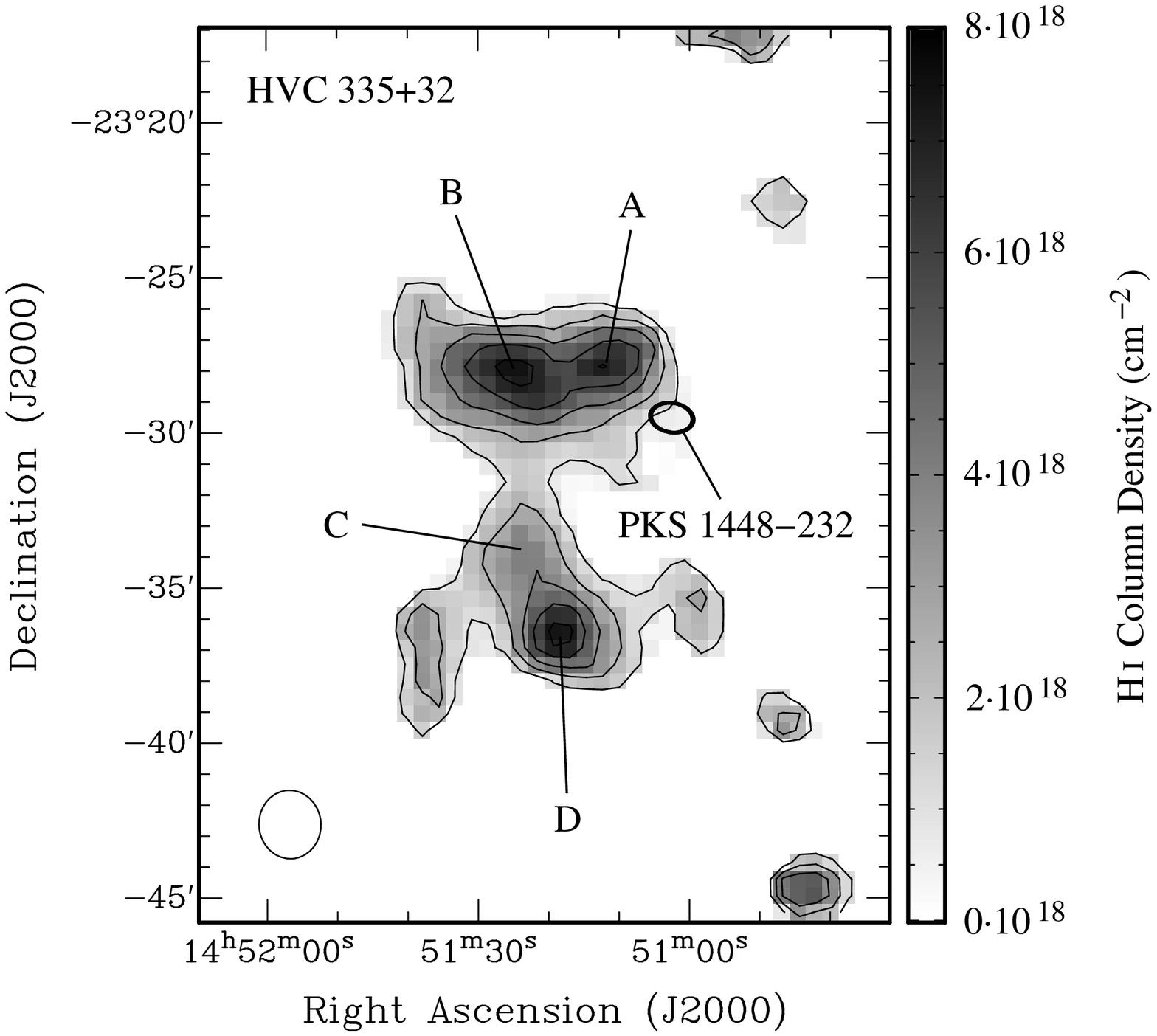}}
\caption[]{
VLA 21cm column-density map of high-velocity H\,{\sc i} gas in the 
direction of the quasar PKS\,1448$-$232 (velocity range: $-130$ to $-170$ 
km\,s$^{-1}$). Several small H\,{\sc i} clumps
with low H\,{\sc i} column densities ($<10^{19}$ cm$^{-2}$) are detected.
The line of sight towards PKS\,1448$-$232 passes the outer envelope 
of clump A. The beam size is indicated in the lower left corner. The contours
refer to H\,{\sc i} column densities of $1,2.5,4,5.5$ and $7\times10^{18}$
cm$^{-2}$.
}
\end{figure}

\begin{table*}
\caption[]{Summary of VLA H\,{\sc i} measurements}
\begin{footnotesize}
\begin{tabular}{crrrrrrr}
\hline
Clump  &  $\alpha$ (J2000)  &  $\delta$ (J2000)  &  $v_{\rm LSR}$  &  FWHM\,$^{\rm a}$    
&  $N$(H\,{\sc i})\,$^{\rm b}$         &  $T_{\rm B}$
&  $T_{\rm kin}^{\rm max}$  \\
&  [h:m:s]           &  [d:m:s]           &  [km\,s$^{-1}$]         &  [km\,s$^{-1}$]  &  [cm$^{-2}$]          &  [K]
&  [K]                      \\
\hline
A      &  14:51:12          &  $-$23:27:45       &  $-147$         &  $13.7\pm0.7$  &  $7.1 \times 10^{18}$  &  $0.34$
       &  $4200$                   \\
B      &  14:51:25          &  $-$23:28:00       &  $-151$         &  $11.3\pm0.6$  &  $7.5 \times 10^{18}$  &  $0.34$
       &  $2900$                   \\
C      &  14:51:23          &  $-$23:34:00       &  $-153$         &   $6.2\pm1.0$  &  $4.0 \times 10^{18}$  &  $0.32$
       &   $900$                   \\
D      &  14:51:18          &  $-$23:36:30       &  $-155$         &  $11.0\pm2.0$  &  $7.4 \times 10^{18}$  &  $0.38$
       &  $2700$                   \\
\hline
\end{tabular}
\\
$^{\rm a}$ FWHM is derived from a Gauss fit.\\
$^{\rm b}$ typical $1\sigma$ error from the fit is $\sim 1.0\times10^{18}$ cm$^{-2}$.
\end{footnotesize}
\end{table*}

\section{Other high-velocity Ca\,{\sc ii} absorbers}

The fact that we have identified a low-column density H\,{\sc i} structure in
Ca\,{\sc ii} and Na\,{\sc i} absorption along a random
line of sight through the halo raises suspicion that such objects have
a significant sky covering fraction and thus may be
ubiquitous in the Milky Way halo. 
To test this idea, we have done a preliminary inspection
of other QSO sight lines through the halo, 
based on high-resolution, high S/N data from the
above mentioned UVES Large Programme.
All QSOs in this sample have Galactic latitudes $|b|>30$.
In eight out of 13 
suitable QSO sightlines 
\footnote{Including the PKS\,1448$-$232 sightline.}
we find high-velocity Ca\,{\sc ii}
absorbers with $|v_{\rm LSR}|=50-200$ km\,s$^{-1}$; many of these
absorbers have no HVC counterparts in the H\,{\sc i} 21cm data of
the LAB survey 
(Kalberla et al.\,2005; Hartmann \& Burton 1997).
Two examples for high-velocity Ca\,{\sc ii} absorption are shown in Fig.\,3.
Many of these sightlines have severe blending problems 
with high-redshift Ly\,$\alpha$ forest lines and intervening metal-line absorbers,
so that a significant fraction of the high-velocity Ca\,{\sc ii} features
in these data even may remain unnoticed. 
However, the number of detected HVC Ca\,{\sc ii} lines
implies that the covering fraction of these low-column density gas clumps 
is large, indeed. All of the detected HVC Ca\,{\sc ii} lines
are narrow (some with multiple velocity components), suggesting the presence
of dense filamentary or clump-like gaseous structures but not large,
extended diffuse clouds.
Associated high-velocity Na\,{\sc i} absorption is seen only
along two sightlines (PKS\,1448$-$232 \& Q\,0109$-$3518). Given the 
coordinates and the observed radial velocities,
many of these low-column density absorbers most likely are associated with 
large, extended HVC complexes such as the Magellanic Stream. 
A more detailed analysis of these Ca\,{\sc ii} absorbers and their 
H\,{\sc i} 21cm counterparts will be presented in a future paper.

\section{Discussion}

Our study suggests that the Milky Way halo contains 
a population of low-column density neutral gas clumps
(see also Hoffman et al.\,2004) that give
rise to weak high-velocity Ca\,{\sc ii} absorption. Due
to the low H\,{\sc i} column densities, many of these
clumps may lie below the detection limit of current
21cm surveys of H\,{\sc i} clouds in the halo.
Many of the high-velocity Ca\,{\sc ii} absorbers probably are associated
with known HVCs, implying that both Galactic and extragalactic
gas contributes to the population of these systems.
Although it remains unclear to what extent such
clouds contribute to the total mass flow of the neutral gas in the
halo, they appear to have a considerable area filling factor
(given the high Ca\,{\sc ii} detection rate).
If low-column density H\,{\sc i} clumps are typical for the halos of spiral
galaxies, they should produce H\,{\sc i} Lyman-Limit absorption
and Mg\,{\sc ii} absorption 
along almost all lines of sight that pass through the 
circumgalactic gas of 
other galaxies. In fact, with typical H\,{\sc i} column densities
$\leq 10^{19}$ cm$^{-2}$ and a complex velocity component structure,
the properties of the Ca\,{\sc ii} 
structures in the Milky Way halo resemble those
of strong Mg\,{\sc ii} absorbers that are nearly 
always found within an impact parameter of $\sim 35\,h^{-1}$ kpc of 
a luminous galaxy (e.g., Ding et al.\,2005; Bergeron \& Boiss\a'e 1991).
These strong Mg\,{\sc ii} systems are believed to sample both disk
and halo gas in galaxies.
The Ca\,{\sc ii} systems presented in this paper possibly
represent the Galactic counterparts of halo Mg\,{\sc ii}
systems with Mg\,{\sc ii} equivalent widths $W_{2796}<1$ \AA.
At H\,{\sc i} column densities $\leq 10^{19}$ cm$^{-2}$ 
the neutral gas fraction probably is only a few percent or less
(Corbelli \& Bandiera 2002), so that one would expect that low-column density
H\,{\sc i} clumps in the halo have substantial ionized gaseous envelopes.
Therefore it seems likely that the detected Ca\,{\sc ii} high-velocity features also are
related to the population of highly-ionized HVCs in the halo seen in C\,{\sc iv} and O\,{\sc vi}
absorption (e.g., Sembach et al.\,2003). 

Summarizing, the observations 
presented in this paper suggest that the weak high-velocity
Ca\,{\sc ii} absorbers may provide an important
link between 
absorption line systems observed in
the circumgalactic environment of other galaxies and the neutral and highly-ionized
high-velocity gas in the halo of the Milky Way.

\begin{figure}[h!]
\resizebox{0.98\hsize}{!}{\includegraphics{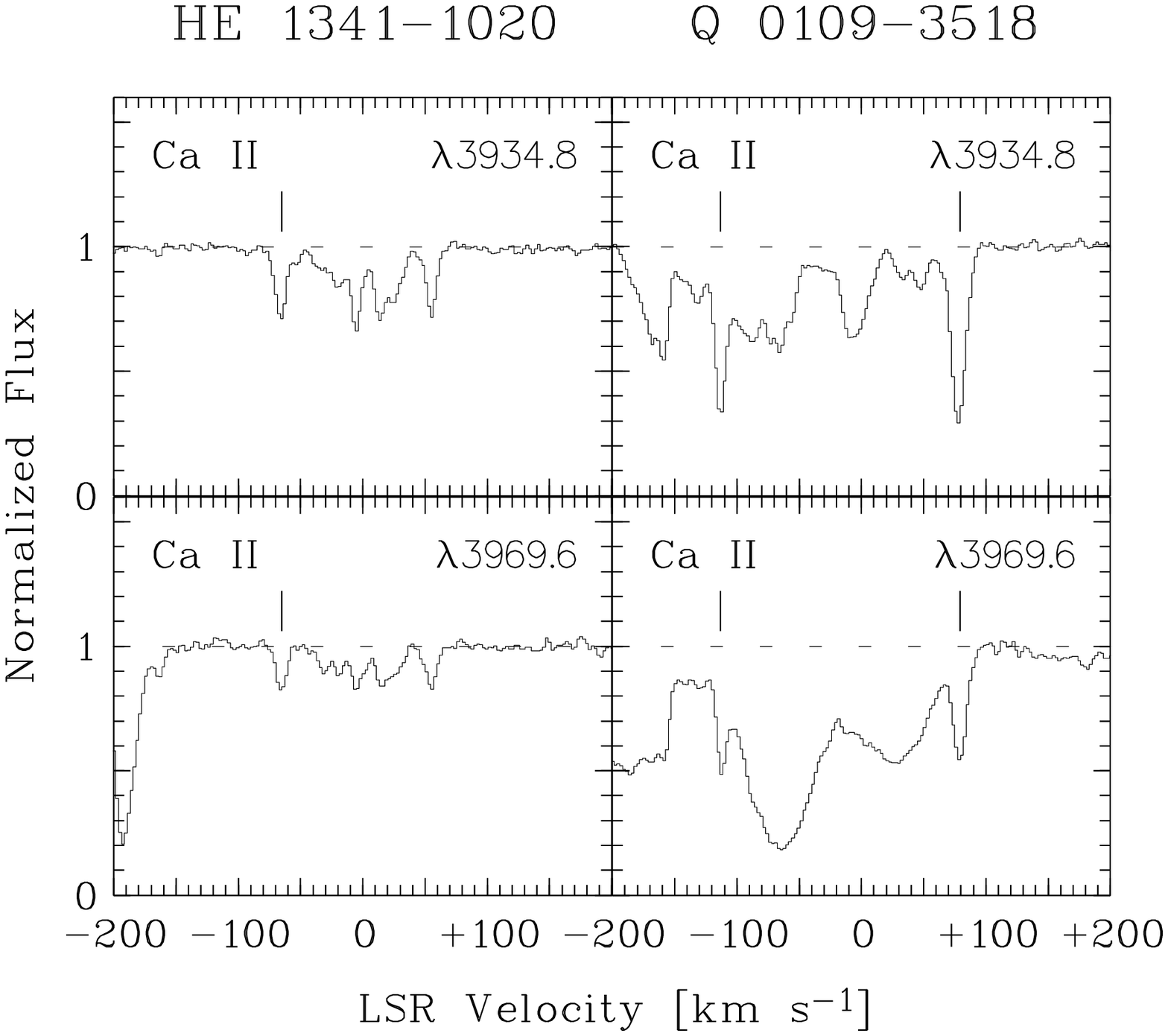}}
\caption[]{High-velocity Ca\,{\sc ii} absorption (indicated
by the long dashes) in the direction of the QSOs HE\,1341$-$1020 ($l=323.5$, $b=+50.2$)
and Q\,0109$-$3518 ($l=275.5$, $b=-81.0$). The absorption near $-70$ km\,s$^{-1}$ 
toward HE\,1341$-$1020 apparently is not related to any known large HVC/IVC complex.
The two absorption
components near $-110$ and $+80$ km\,s$^{-1}$ toward Q\,0109$-$3518 most likely
are associated with the Magellanic Stream. H\,{\sc i} emission at 
$v_{\rm LSR}=-100$ to $-180$ km\,s$^{-1}$ is seen in the LAB survey in the immediate
environment of this sightline.
}
\end{figure}

\begin{acknowledgements}

P.R. acknowledges financial support by the German
\emph{Deut\-sche For\-schungs\-ge\-mein\-schaft}, DFG,
through Emmy-Noether grant Ri 1124/3-1. T.W. is supported
by the DFG through grant KE 757/4-1. We thank K.S. de\,Boer, 
J. Bergeron \& B.P. Wakker for helpful comments.

\end{acknowledgements}

\end{document}